\documentclass[preprint,longbibliography,superscriptaddress]{achemso}
\usepackage{graphicx}%
\usepackage[T1]{fontenc}
\usepackage{xcolor}
\usepackage{amsmath}
\usepackage{upgreek}
\title{Valley-Polarized Exciton Dynamics in Exfoliated Monolayer WSe$_2$}
\author{Gerd Plechinger}
\email{gerd.plechinger@physik.uni-r.de}
\author{Tobias Korn}
\author{John M. Lupton}
\affiliation{Institut f\"ur Experimentelle und Angewandte Physik,
	Universit\"at Regensburg, D-93040 Regensburg, Germany}
\begin{document}
\begin{abstract}
Semiconducting transition metal dichalcogenide monolayers have emerged as promising candidates for future valleytronics-based quantum information technologies. Two distinct momentum-states of tightly-bound electron-hole pairs in these materials can be deterministically initialized via irradiation with circularly polarized light. Here, we investigate the ultrafast dynamics of such a valley polarization in monolayer tungsten diselenide by means of time-resolved Kerr reflectometry. The observed Kerr signal in our sample stems exclusively from charge-neutral excitons. Our findings support the picture of a fast decay of the valley polarization of bright excitons due to radiative recombination, intra-conduction-band spin-flip transitions, intervalley-scattering processes, and the formation of long-lived valley-polarized dark states.
\end{abstract}
\maketitle
\footnote{This document is the unedited Author's version of a Submitted Work that was subsequently accepted for publication in The Journal of Physical Chemistry C, copyright $\copyright$ American Chemical Society after peer review. To access the final edited and published work see \\ http://pubs.acs.org/articlesonrequest/AOR-hCJpKkDiPY2E8AZzPe3t .}
\section{Introduction}
 The family of layered crystalline semiconductors, in particular the transition metal dichalcogenides (TMDCs), has attracted vast attention in the physical and chemical sciences during the last few years. This interest can be traced back to their particular material properties: thinned down to a single monolayer unit, which is only three atoms thick, these materials become direct semiconductors with optical bandgaps in the visible and near-infrared spectral range located at the K points of the Brillouin zone \cite[]{Splendiani2010, Mak2010, Tonndorf2013}. The TMDCs can thus be considered as gapped graphene analogues. Additionally, due to a spatial and dielectric confinement into two dimensions, the Coulomb interaction between electrons and holes is strongly enhanced, resulting in the formation of excitons with anomalously high binding energies in the range of several hundred meV \cite[]{Klots2014, Chernikov2014, Ugeda2014, He2014, Hanbicki2015}. As the TMDC crystals have a diatomic basis, the six K valleys located at the edges of the hexagonal Brillouin zone can be divided into two different groups: +K and -K, lying at opposite corners in the reciprocal space. The broken inversion symmetry and the symmetries of the conduction and valence bands in a monolayer lead to different selection rules for left- ($\sigma^-$) and right-handed circularly polarized light ($\sigma^+$): while $\sigma^-$-polarized photons are absorbed exclusively in the -K valley, the +K valley is only sensitive to $\sigma^+$-polarized photons \cite{Xiao2012}. Such a valley polarization can be read out by polarization-resolved photoluminescence experiments \cite{Mak2012, Zeng2012, Sallen2012} and exhibits considerable potential for the application in future valley-based quantum information technologies \cite{Xu2014, Schmidt2016}. Upon optical band-to-band excitation, the valley pseudospin is not imprinted on single charge carriers but on excitonic quasiparticles, adding further complexity to the system. In tungsten-based TMDCs, the spin splitting in the conduction band is predicted to be negative \cite{Kormanyos2015}, resulting in the possibility of the existence of optically dark excitonic levels having lower energy than the optically addressable ones \cite{Plechinger2016a, Zhang2015b, Zhang2016, Molas2016}. A concrete understanding of the valley dynamics in these complex excitonic systems is of utmost importance as a starting point for engineering ultra-long valley lifetimes in TMDCs in order to potentially make valley physics the foundation for a competitive quantum information technology.

 Here, we present time-resolved Kerr rotation (TRKR) measurements which provide insight into the valley dynamics of excitons in monolayer WSe$_2$ at liquid-helium temperatures. By tuning the excitation energy, we observe an excitonic resonance in the TRKR signal at the exact photon energy of the charge-neutral exciton photoluminescence. The valley dynamics arise from two subpopulations: a very fast decaying bright exciton population within the first few picoseconds, and a slowly decaying population of either dark excitons or resident carriers, supporting the suggestion of negative conduction band spin-splitting in WSe$_2$ \cite{Kormanyos2015}.

 \section{Results and Discussion}

 \begin{figure}
 	\includegraphics*[width=\textwidth]{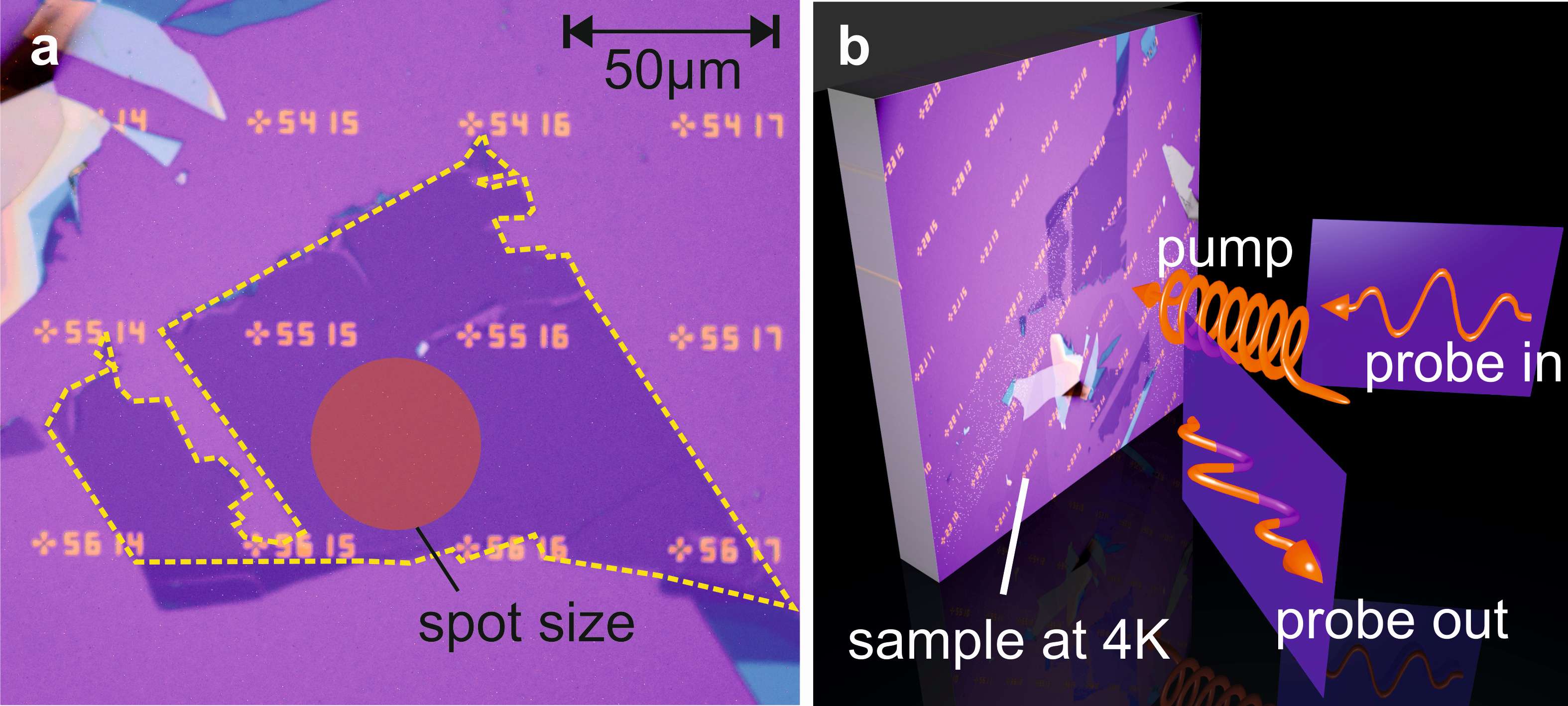}
 	\caption{(a) Optical microscope image of the investigated monolayer WSe$_2$ flake. The red circle indicates the laser spot size. The numbers are position markers. The perimeter of the monolayer region of the flake is marked in yellow. (b) Schematic illustration of the time-resolved Kerr reflectometry: first, a circularly polarized pump pulse is focussed on the sample at cryogenic temperatures in order to create a valley polarization, then a linearly polarized probe pulse is directed onto the same measurement spot. The polarization axis of the latter is tilted due to the magneto-optical Kerr effect.}
 	\label{fig:sample}
 \end{figure}

 Figure \ref{fig:sample}(a) shows a microscope image of the investigated monolayer WSe$_2$ flake produced by mechanical exfoliation and subsequent all-dry transfer onto a silicon wafer piece with 285\,nm thermal oxide and gold markers on top. The sample size is large enough to guarantee that the laser spot (diameter $\sim 30$\,$\mu$m) lies fully within the area of the monolayer. A schematic of the TRKR technique is shown in Fig. \ref{fig:sample}(b): a circularly polarized pump pulse is focussed onto the TMDC flake to generate a valley polarization. The flake is visualised with a digital microscope system. After a variable time delay $\Delta t$, a linearly polarized probe pulse is directed onto the same measurement spot. The optically generated valley polarization in the sample leads to a slight rotation of the polarization axis of the probe beam. By varying the time delay $\Delta t$, we receive a stroboscopic time trace of the overall valley polarization.

 \begin{figure}
 	\includegraphics*[width=\textwidth]{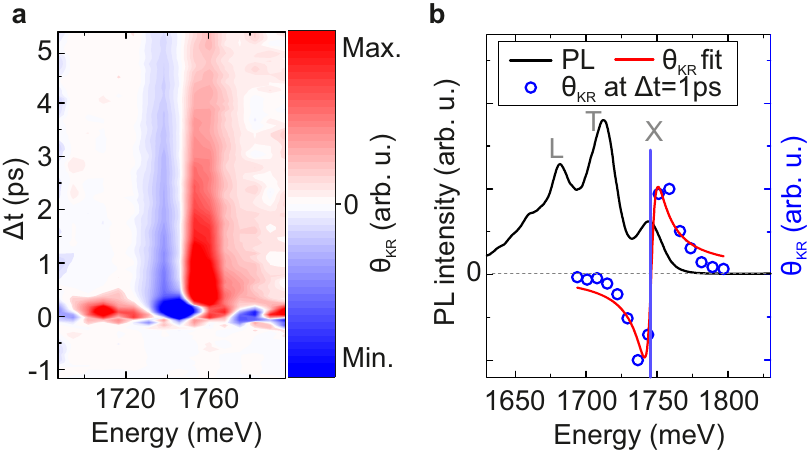}
 	\caption{(a) False-color plot of the Kerr-rotation angle $\theta_{KR}$ as a function of the degenerate excitation and detection laser energy and the time delay $\Delta t$ between pump and probe pulse. (b) Direct comparison of the photoluminescence spectrum at 4\,K (black line) and $\theta_{KR}$ at a fixed time delay of $\Delta t=1$\,ps plotted in dependence of the laser energy (blue circles). The blue vertical line indicates the exciton resonance energy extracted from the photoluminescence spectrum, the red line a dispersive Lorentzian fit to the Kerr data.}
 	\label{fig:lambda}
 \end{figure}

 First, we discuss excitation-energy dependent TRKR measurements. Supplementary Figure 1 shows the temporal evolution of the Kerr rotation angle for left-hand circularly polarized, right-hand circularly polarized as well as for linearly polarized excitation. No significant signal is observed for linearly polarized excitation emphasizing that the Kerr rotation is based exclusively on the valley-selective absorption of circularly polarized light. Figure \ref{fig:lambda}(a) displays the Kerr rotation angle $\theta_{KR}$ on a monolayer of WSe$_2$ at a temperature of 4\,K on a false-color scale as a function of the photon energy of the degenerate pump and probe pulses and of the delay time $\Delta t$ between both pulses. Note that $\theta_{KR}$ can be positive or negative. At $\Delta t=0$\,ps, pump and probe beam arrive at the same time on the sample and generate a large Kerr signal. The large signal variations at zero time delay arise mainly from interference effects between the two pulses. In the range between 1730\,meV and 1770\,meV, a clearly discernible decay behaviour prevails in the Kerr signal at later times. Furthermore, a sign change of $\theta_{KR}$ is observed at 1745\,meV over the entire time range investigated. This behaviour is seen more clearly in Fig. \ref{fig:lambda}(b), where a slice through the colour map in (a) at $\Delta t=1\,$ps is shown. The $\theta_{KR}$ curve resembles a dispersive Lorentzian function, providing evidence that the underlying signal originates from a Lorentzian-type excitonic resonance \cite{Fokina2010, Glazov2012, Lee2015, Plechinger2016a}. Theoretically, in the absence of substrate-induced interference effects \cite{Chen2005}, $\theta_{KR}$ as a function of the excitation energy $E$ should follow the relation $\theta_{KR} \propto \frac{(N^+ - N^-)(E_0-E)}{(E_0-E)^2+(\Gamma)^2}$, with $N^{\pm}$ being the exciton populations in +K and -K valleys, $E_0$ the resonance energy, and $\Gamma$ the damping rate \cite{Fokina2010, Glazov2012, Lee2015, Plechinger2016a}. The experimental data can be accurately modelled with the above proportionality, as depicted by the red curve in Fig. \ref{fig:lambda}(b). Next, we compare the resonance energy $E_0$ of 1745\,meV extracted from the TRKR measurements with the energies of the optical emission features. It is known that the Stokes shift between absorption and emission in 2D TMDCs is negligible \cite{Zhao2013a}. Therefore, we can use the PL resonance energy as a direct measure for the absorption resonance energy which is in turn fully compatible with the observed Kerr resonance. Figure \ref{fig:lambda}(b) displays a normalized low-temperature photoluminescence spectrum of monolayer WSe$_2$. The different photoluminescence peaks arise from the radiative recombination of separate excitonic quasiparticles: localized excitons (L) emitting at around 1682\,meV, trions (T) at 1713\,meV and charge-neutral excitons (X) at 1745\,meV \cite{Jones2013, Wang2014c, Arora2015}. The photoluminescence peak energy for the X species matches the TRKR resonance energy. Around 1720\,meV, only a very weak TRKR signal is observed, which is fully reproducible with a model in which only one single exciton resonance (X) is considered. We therefore conclude that, when comparing the TRKR data and the PL spectrum, no significant Kerr signal stemming from trions or localized excitons can be identified. The absence of a trion resonance in TRKR could indicate only moderate natural doping of the investigated TMDC crystal. Moreover, singlet and triplet trion species \cite{Jones2016} are expected to have proportionality factors of opposite sign to determine the Kerr resonance curves \cite{Plechinger2016a}, which might lead to a partial cancellation of the TRKR signal.

\begin{figure}
	\includegraphics*{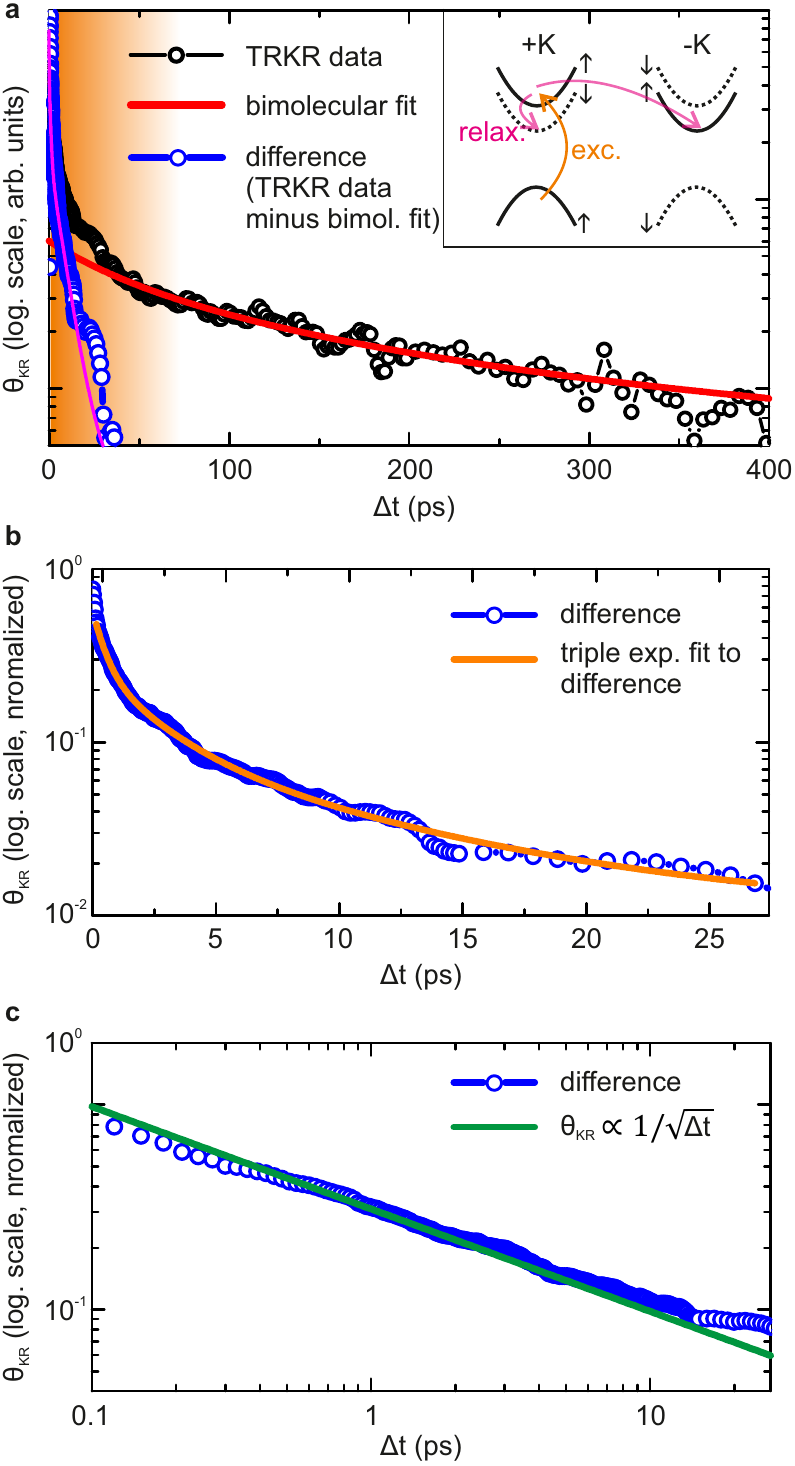}
	\caption{(a) Kerr angle $\theta_{KR}$ as a function of the delay time $\Delta t$ between pump and probe pulse. The red curve represents a bimolecular fit of the long-lived part of the signal, the blue curve is the difference between the measurement data and the bimolecular decay. The inset displays a sketch of the bandstructure at the +K and -K points. The orange arrow symbolizes the allowed A exciton transition in the +K valley, the violet arrows indicate intra- and intervalley relaxation of electrons in the conduction band. (b) Plot of the blue difference data set for the first few picoseconds. The orange line shows a triple exponential fit. (c) Double-logarithmic representation of the difference data. The green curve shows a power-law with an exponent of $-0.5$.}
	\label{fig:time}
\end{figure}

In the following, the excitation and detection energy is kept fixed at 1730\,meV, the energy of the strongest Kerr signal. We now turn to the dynamics of the valley polarized excitons as shown in Fig. \ref{fig:time}(a). For small delay times $\Delta t<25$\,ps, one observes a rapid decay of the TRKR signal over more than one order of magnitude. As our measurement method is sensitive to the total valley polarization and not only to the degree of polarization, a decay of the overall exciton population is also observable in TRKR. The reported ultrafast radiative recombination of the excitons in WSe$_2$ on the time scale of 150\,fs \cite{Poellmann2015} is therefore likely to be, in part, responsible for the fast initial decay of the Kerr signal. Additionally, theory predicts a very fast valley depolarization on the timescale of a few picoseconds due to intervalley electron-hole exchange interactions \cite{Yu2014a, Zhu2014b, Glazov2014}. However, as the efficiency of this mechanism depends on the size of the splitting of the bright exciton dispersion, which in turn increases proportionally with the exciton momentum, valley depolarization due to the exchange interaction is expected to be of minor importance for resonant excitation and the low excitation densities on the order of 2\,kWcm$^{-2}$ used in our experiment. Nevertheless, this mechanism could represent another decay channel observable in the TRKR time trace. Trion formation might also act as a signal quencher for charge-neutral excitons, because the trions themselves, even in the case when they are valley polarized, do not lead to a significant Kerr rotation of the probe pulse in WSe$_2$ (cf. Fig. \ref{fig:lambda}(b)) - in contrast to WS$_2$ \cite{Plechinger2016a} - and thus reduce the measurable Kerr signal.

For tungsten-based TMDCs, fast intra-conduction-band transitions of electrons are very likely. The reason for this is the negative conduction-band spin-splitting. In an A exciton in WSe$_2$, the bound electron and hole are located in the upper valence band and in the upper conduction band \cite{Kormanyos2015}. For the electron, this is no favorable state. An intra-valley spin-flip or an intervalley spin-conserving relaxation to the lower-lying conduction subband would lead to an overall energy reduction for the exciton (see inset in Fig. \ref{fig:time}(a)). In the picture of bound excitons, this scenario would correspond to an additional excitonic state at lower energies than the optically addressable one. Here, the Coulombic exchange interaction is of major importance and adds to the single-particle conduction band splitting, resulting in a theoretical exciton splitting on the order of 20\,meV for WSe$_2$ \cite{Echeverry2016}. The lower-lying exciton level cannot be excited optically and cannot decay radiatively, and hence represents a dark excitonic state. This becomes clearer when looking at the single-particle picture: the symmetry-dictated selection rules in monolayer TMDCs only allow transitions between bands having the same spin orientation, that is, from the lower valence subband to the lower conduction subband (B exciton transitions) and from the upper valence subband to the upper conduction subband (A exciton transitions). Radiative recombination from states with the electron in the lower conduction subband and the hole in the upper valence subband is dipole forbidden. As our experiment probes, among other things, the population of the upper valence subband and the upper conduction subband, a depopulation of the upper conduction subband leads to a partial reduction of the measureable Kerr signal. Given that the Fermi energy does not lie within the bright-exciton level, because of the low sample temperature, the low excitation power, and the moderate natural doping of the sample, these bright-to-dark excitonic transitions may turn out to be responsible for an ancillary fast decay of our Kerr rotation signal within the first few ps.

As a consequence, a significant population of dark excitons is created. As the presence of valley-polarized holes in the upper valence subband selectively blocks in part the absorption of the probe beam, dark excitons can be observed in TRKR. The slow signal decay for $\Delta t>50$\,ps, shown in Fig. \ref{fig:time}(a), presumably stems from these dark excitons. These can only recombine nonradiatively via exciton-exciton annihilation. Such a process manifests itself in bimolecular dynamics described by the quadratic rate equation

\begin{equation} \frac{\partial N_X}{\partial t} = a \cdot N_X^2 , \end{equation}

with the exciton density $N_X$ and a proportionality constant $a$. We can fit the long-lived tail of the Kerr signal with the above relation (red curve in Fig. \ref{fig:time}(a)). We note that an alternative explanation for the measured signal at delay times exceeding 100\,ps would be the transfer of spin polarization to resident carriers, as suggested in similar measurements on MoS$_2$ \cite{Yang2015c}, WS$_2$ \cite{Yang2015a} and also WSe$_2$ monolayers grown by chemical-vapor deposition \cite{Hsu2015, Song2016, Bushong2016}, where nanosecond spin lifetimes are reported. Here, however, in contrast to these previous studies, we investigate mechanically exfoliated, undoped samples and the signal decay occurs on the sub-nanosecond time scale. We cannot discriminate whether we observe valley-polarized dark excitons or spin-valley-polarized electrons in the Kerr dynamics. However, as can be seen from the false-colour plot in Fig. \ref{fig:lambda}(a), the Kerr signal is antisymmetric in energy with a time-independent resonance energy of 1745\,meV, at least within the first 5\,ps. The Kerr signal clearly indicates an antisymmetric Lorentzian-shaped resonance, as would be expected for excitonic quasiparticles. This observation is in striking contrast to the symmetric Kerr signal observed for resident carriers in MoS$_2$ \cite{Yang2015c}.

Subtracting the bimolecular fit from the measurement data in the first 30\,ps, we end up with a decay curve as depicted in Fig. \ref{fig:time}(b). A triple exponential decay reproduces the time-trace accurately, implying a clear deviation from a monoexponential decay that would hint at one single depolarization mechanism. Instead, we have three different contributions to the fast initial signal reduction with decay times of 500\,fs, 2\,ps and 10\,ps, respectively. The fastest decay of 500\,fs be attributed to the radiative recombination of the excitons, where the timescale is in line with previous reports \cite{Poellmann2015, Robert2016}. The origin of the 2\,ps and 10\,ps decays may tentatively be found in the bright-to-dark exciton transitions, the trion formation processes, or valley depolarization due to exchange interactions.

We note that the valley-dynamics presented in Fig. \ref{fig:time}(b) can also be analysed in double-logarithmic representation, as shown in Fig. \ref{fig:time}(c). The temporal evolution of the number of valley-polarized bright excitons appears to closely follow a power-law described by the relation $\theta_{KR} \propto 1/\sqrt{t}$, shown as the green line. We refrain from speculating on which of the two representations of the data is most appropriate.

At present, we can only conclusively deduce the presence of additional dynamics in the TRKR signal and hence in the valley-polarized exciton population beyond radiative exciton recombination. Discrimination between these additional mechanisms will necessitate TRKR measurements at non-zero magnetic fields, as well as a control of the carrier density by deliberate doping or electrostatic gating.

\section{Conclusion}

In conclusion, we have performed TRKR measurements on a mechanically exfoliated monolayer of WSe$_2$. Tuning of the degenerate excitation and detection photon energy reveals a single excitonic resonance at 1745\,meV, which can be identified as the charge-neutral exciton upon comparison with the photoluminescence spectrum. The Kerr trace exhibits a fast initial decay within the first few picoseconds, and a subsequent slow decay with a half-life of 170\,ps. This result is in line with a picture where the initial fast signal loss stems, amongst others, from radiative recombination and bright-to-dark exciton transitions. The long-lived part of the Kerr transient is tentatively assigned to exciton-exciton annihilation in a dark-exciton subpopulation. This interpretation further strengthens the assumption of a negative conduction-band splitting in tungsten-based TMDCs \cite{Kormanyos2015}. The ultrafast intrinsic exciton dynamics appear to constitute the major limiting factor for the valley lifetime of bright excitons in mechanically exfoliated WSe$_2$. At the same time, optically dark states - either dark excitons or resident carriers - can maintain the spin-valley polarization for a longer time period than bright excitons and could therefore be the focus of future studies with the prospect of realising information encoding through long-lived valley-pseudospin polarization. The measurements presented here provide further insight into valley dynamics in this fascinating class of materials and constitute a contribution on the way to future valleytronics technologies.

\section*{Methods}
\subsection*{Sample preparation}
Monolayer flakes of WSe$_2$ were mechanically exfoliated from bulk crystals (hq Graphene inc.) onto Gel-Film$^{\textregistered}$ substrates from Gelpak. Through an all-dry transfer technique \cite{Castellanos2014}, the flakes were placed on a Si chip with 285\,nm thermal SiO$_2$ and gold markers on top.
\subsection*{Time-resolved Kerr reflectometry}
The pulses of a frequency-tunable Ti:Sapphire laser (Chameleon Ultra II, Coherent) with 130\,fs pulse length are split into pump and probe pulses. The pump pulses are circularly polarized with a linear polarizer and a quarter-wave plate and focussed onto the sample with a 75\,mm achromatic lens. The probe pulses are guided to a delay stage, where a variable time delay between pump and probe pulse can be generated. Subsequently, they are linearly polarized and also focussed onto the sample. The polarization rotation of the reflected probe beam is analyzed with a Wollaston prism, two Si photodiodes and a lock-in amplifier.

\section*{Additional Information}
\subsection{Supporting Information}
Kerr rotation measurements on monolayer WSe$_2$ flake using circularly and linearly polarized excitation.
\subsection*{Acknowledgements}
The authors gratefully acknowledge financial support by the DFG through SFB689, KO3612/1-1, and GRK1570.
\subsection*{Author contributions}
G.P., T.K., and J.M.L. conceived the experiment. G.P. performed the experiments. G.P. and T.K. analysed the data. G.P. wrote the manuscript with input from all authors.
\subsection*{Competing financial interests}
The authors declare no competing financial interests.
\providecommand{\latin}[1]{#1}
\providecommand*\mcitethebibliography{\thebibliography}
\csname @ifundefined\endcsname{endmcitethebibliography}
  {\let\endmcitethebibliography\endthebibliography}{}

\section{TOC Graphic}
\begin{figure}
	\includegraphics*{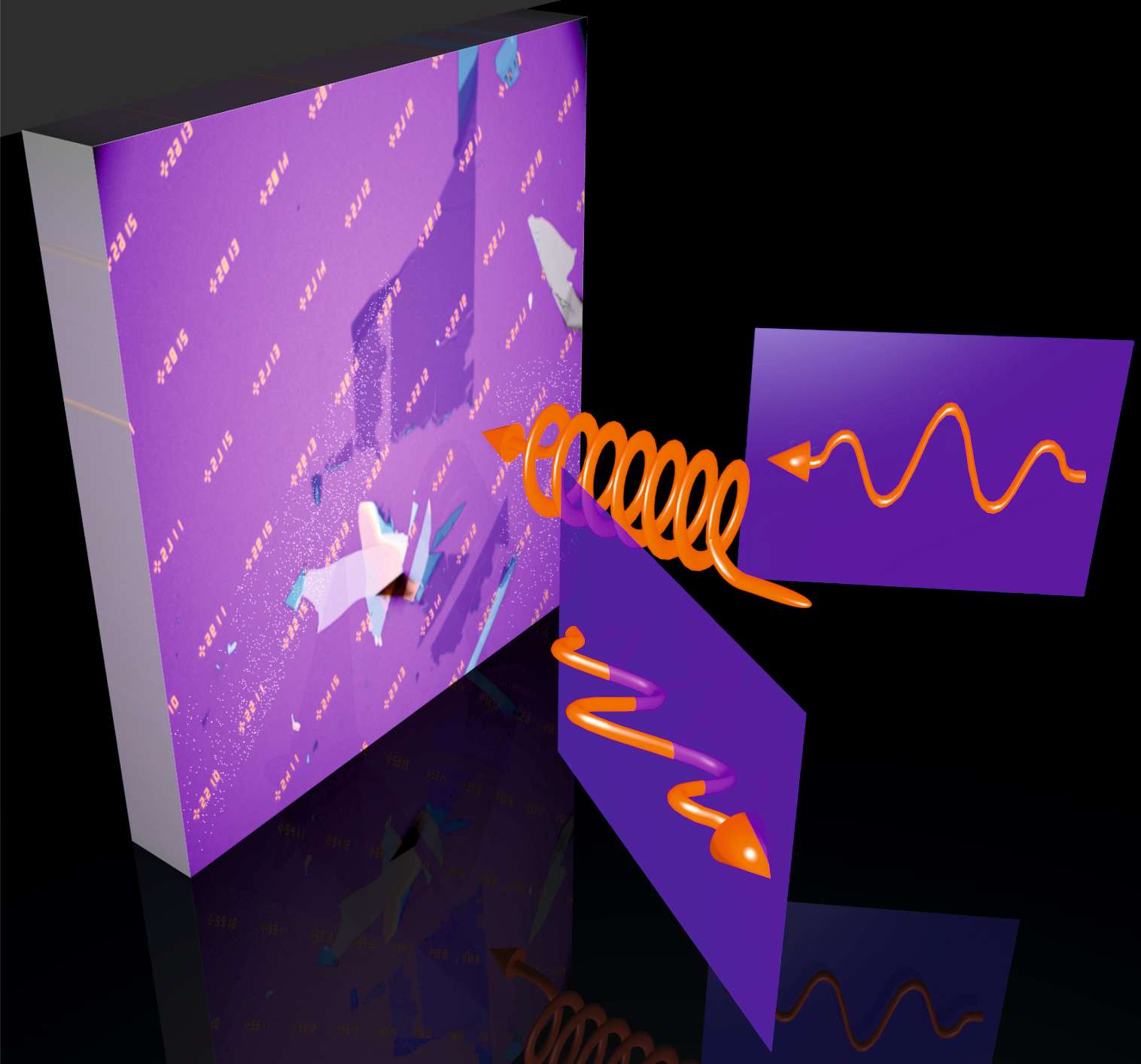}
\end{figure}

\end{document}